\documentclass{aa}
\usepackage{natbib} 
\usepackage{amssymb}
\usepackage{amsmath}
\usepackage{graphicx} 
\usepackage{txfonts}

\def\msun{${\rm M}_{\odot}$}

\def\lsun{$L_{\odot}$}

\newcommand{\chem}[2]{$\rm{}^{#1}\kern-0.8pt#2$}
\newcommand{\chim}[2]{\rm{}^{#1}\kern-0.8pt#2}
\newcommand{\reac}[6]{$\rm\,{}^{#1}\kern-0.8pt{#2}\,({#3}\,,{#4})\,{}^{#5}\rm{#6}\,$}

\begin{document} 
\title{V453 Oph: a s-process enriched, but carbon-deficient RV Tauri
star of low intrinsic metallicity \thanks{based on observations
collected at the European Southern Observatory (La Silla) in Chile
(67.D-0054(A) and on the 1.2m Swiss Euler telescope also on La Silla}
}
 
\author{Pieter Deroo\inst{1} \and Maarten Reyniers\inst{1} \and Hans
Van Winckel\inst{1} \and St\'ephane Goriely\inst{2} \and Lionel
Siess\inst{2}}
 
\institute{Instituut voor Sterrenkunde, K.U. Leuven, Celestijnenlaan
200B, B-3001 Leuven, Belgium \and Institut d'Astronomie et
d'Astrophysique, Universit\'e Libre de Bruxelles, CP 226, 1050
Brussels, Belgium}
 
\date{Received <date> / Accepted <date>} 
\offprints{P. Deroo \\ \email{Pieter.Deroo@ster.kuleuven.ac.be}}

\abstract{ This paper reports the detection of a heavy element
enriched RV\,Tauri variable with an abundance pattern that differs
significantly from a standard s-process enriched object: \object{V453
Oph}. Based on optical high-resolution spectra, we determined that
this object of low intrinsic metallicity ([Fe/H]$=-2.2$) has a mild,
but significant, enrichment ([s/Fe]$\sim$+$0.5$) of heavy elements for
which the distribution points to slow neutron capture nucleosynthesis.
This result is strengthened by a comparative analysis to the
non-enriched RV\,Tauri star \object{DS Aqr} ([s/Fe]$=$ 0.0). Although
V453\,Oph is the first RV\,Tauri star showing a strong s-process
signature, it is \emph{not} accompanied by C enhancement,
challenging our current nucleosynthetic models of post-AGB stars that
predict a simultaneous enrichment in C and s-process elements. The low
N abundance excludes CN cycling as being responsible for the low C
abundance. We explore three different scenarios to explain the heavy
element distribution in this evolved object: an enrichment of the
parental cloud, an accretion scenario in which the chemical patterns
were acquired by mass transfer in a binary system and an intrinsic
enrichment by dredge-up. 

 } 
\titlerunning{V453 Oph, s-process enriched but C-deficient } 
\authorrunning{P. Deroo et al.}  
\maketitle

\section{Introduction} 
 
RV\,Tauri stars are rare supergiants that occupy the high-luminosity
end of the pop. II Cepheid instability strip. The identification of
RV\,Tauri stars is based on their characteristic light curves, which
show alternating deep and shallow minima with periods ranging from 30
to 150 days. This variability is often accompanied with lower
amplitude irregularities and cycle-to-cycle variability
\citep[e.g. review on pop. II Cepheids by][]{wallerstein02}.
  
The far-IR excess of RV\,Tauri stars due to circumstellar dust was
detected by IRAS and led \cite{jura86} to conclude that they are young
post-AGB stars, in which the circumstellar dust is a relic of the
strong dusty mass loss on the AGB. For many RV\,Tauri variables, a
near-IR excess is detected as well \citep{Gehrz_1972,
lloydevans85}. The genuine high-luminosity nature of the RV\,Tauri
class was confirmed by the detection of some members in Globular
Clusters and in the LMC. The LMC variables obey a period-luminosity
(P-L) relation, which follows the high luminosity extension of the P-L
relation of W\,Vir stars \citep{alcock98}.
  
Chemically, however, RV\,Tauri stars do not show evidence for
post-third dredge-up enrichment: no high C-abundances either
s-process overabundances due to a third dredge-up have been found. A
summary of the available quantitative abundance data on RV\,Tauri
stars can be found in \cite{Giridhar_2000} and \cite{Maas_2004}. Many
RV\,Tauri stars show evidence of selective depletion in which the
refractory elements are depleted relative to the non-refractory
ones. These photospheric chemical patterns are the result of a poorly
understood process in which the circumstellar dust is separated from
the circumstellar gas, followed by a selective re-accretion of the
non-refractories \citep{Lamers_1992, Waters_1992}. Important
diagnostics for this process are high [Zn/Fe], [Zn/Ti] and
[S/$\alpha$] ratios, because Zn and S have a very low condensation
temperature with respect to Fe, Ti and the $\alpha$-elements
. The process will also result in a C/O ratio larger than
solar due to the stability of the CO molecule, which is thought to be
abundant in the gas accreted onto the stellar surface. Since s-process
elements are refractory, a subsequent depletion process can mask an
intrinsic AGB enrichment. Nevertheless, overabundances of s-process
elements relative to other species with similar condensation
temperature were undetected.
  
The depletion process can be very efficient in post-AGB binary stars
for which there is evidence that the dust is trapped in a
circumstellar disc \citep{vanwinckel95}. This geometry may well be a
necessary condition for the depletion process to be efficient
\citep{Waters_1992}. Evidence is growing that also in dusty RV\,Tauri
stars the dust is trapped in a stable disc
\citep{lloydevans85,vanwinckel99} and the binary nature of several
dusty RV\,Tauri stars has been established \cite[][and references
therein]{vanwinckel03}. It is, however, not generally accepted that
all dusty RV\,Tauri stars reside in binaries. There is no clear
relation between the current IR-excess and the strength of the
depletion process \citep{Giridhar_2000}.
 
The traditional spectroscopic classification of RV\,Tauri stars by
\cite{Preston_1963} is strongly determined by the efficiency of the
depletion process. Spectroscopic group A consists of stars of spectral
type G or K, showing strong absorption lines but normal CN or CH
bands, while some also show TiO bands at minimum light. Group B stars,
being generally somewhat hotter, show weak metal lines combined with
enhanced CN and CH bands. Members of group C resemble those of group
B, except that enhanced bands of CN are not seen. It is now clear that
the low metallicity of the weak-lined group B stars is determined by
the efficiency of the depletion process. Depletion patterns are less
strong, if present at all, in group A stars.  Group C stars are not
affected by depletion, probably due to their low intrinsic metallicity
\citep{Giridhar_2000}.

In this paper we focus on the chemical analysis of two RV\,Tauri
stars, \object{V453\,Oph} and \object{DS\,Aqr} of spectroscopic class
C to study potential AGB enrichment. The
luminosity of these stars is about 2400 $\pm$ 1600\,L$_\odot$, as
determined using the P-L relation of RV\,Tauri variables in the LMC
\citep{alcock98}. This high luminosity identifies them as likely
post-AGB stars \citep[e.g.][]{Blocker_1995}. Since class C objects are
not affected by the depletion process, these intrinsically metal-poor
objects are ideal to study AGB enrichment and possibly the
metallicity dependence of the s-process. Some basic data of both stars
are shown in Table \ref{basic_data}.
 
\begin{table}[t] 
\centering 
\caption{Some basic parameters of V453 Oph and DS Aqr. The pulsation
periods, spectral types and magnitude ranges are from
\cite{Kholopov_1998}. The other parameters were obtained from
Simbad.}\label{basic_data}
\begin{tabular}{lcrr} 
\multicolumn{1}{c}{Data} & & \multicolumn{1}{c}{V453 Oph} &
\multicolumn{1}{c}{DS Aqr}\\
 
\hline \hline
 
Other name: & & BD-02 4345 & HD 216457 \\ Coordinates: &
$\alpha_{2000}$ & +17 26 49.13 & +22 53 17.04 \\ & $\delta_{2000}$ &
-02 23 36.44 & -18 35 30.99 \\ Min. mag.: & & 10.4\,mag & 11.08\,mag
\\ Max. mag.: & & 11.53\,mag & 11.90\,mag \\ Spectral type: & & Fp &
F2II \\ Period: & & 81.3\,d & 78.213\,d \\ Galactic & l & 20.72 &
44.10 \\ coordinates: & b & 17.57 & -61.56 \\
\end{tabular} 
 
\end{table}

After an overview of the observations, we discuss our spectroscopic
parameter determination in Sect. \ref{parameter_determination}. The
abundance pattern for \object{V453\,Oph} and \object{DS\,Aqr} is
derived in Sect. \ref{sect_abundance} and discussed in
Sect. \ref{sect_discussion}. In Sects. 6 and 7 we explore different
scenarios to explain the observed abundances. We end with the
conclusions in Sect.\,\ref{sect_conclusion}.
 
\section{Observations} 
 
\subsection{Spectroscopic}\label{observations_spectroscopic} 
High-resolution spectra of both RV\,Tauri variables were acquired with
the fibre fed spectrograph FEROS, mounted on the $1.5$\,m ESO
telescope at La Silla. The resolving power of the spectra is R $\sim$
48000 at 5000\,\AA\, and a wavelength coverage ranging from
3600\,\AA\, to 9200\,\AA\, is obtained in a single exposure. Two exposures
were obtained for \object{V453\,Oph} and one for DS Aqr. The log of
the observations is reported in Table \ref{basic_data_spectra}.
 
\begin{table}[h]  
\centering 
\caption{The log of the obtained FEROS spectra together with the
signal-to-noise (S/N) ratio in different spectral regions. The S/N
ratio listed for \object{V453\,Oph} corresponds to the final
spectrum.}\label{basic_data_spectra}
 
\begin{tabular}{llcccc} 
\multicolumn{1}{c}{Name} & Date and UT & S/N & S/N & S/N\\ 
& & 400\,nm & 585\,nm & 770\,nm \\ \hline \hline 
V453 Oph &  28/06/01 05:30 & & & \\ 
& 28/06/01 06:25  \raisebox{1.5ex}[0pt]{\bigg\rbrace}  &
 \raisebox{1.5ex}[0pt]{50}
 &\raisebox{1.5ex}[0pt]{180}
 &\raisebox{1.5ex}[0pt]{180}\\
 DS Aqr & 27/06/01 09:05 & 50 & 110 & 80 \\
 
\end{tabular} 
\end{table} 
 
The reduction of the spectra was performed in the standard fashion,
using the FEROS DRS package in the MIDAS environment. Our reduction
includes bias correction, background subtraction, cosmic hit cleaning,
flatfielding, order extraction and wavelength calibration. Ultimately,
the spectra of the different orders were merged and normalised.
 
For V453 Oph, the two normalised spectra were averaged, weighted by
their signal-to-noise (S/N) ratio. In Table \ref{basic_data_spectra}
the S/N ratio is given for different spectral regions.  Sample spectra
are shown in Figs. \ref{fig:nsynth}, \ref{spectrum_synthese} and
\ref{fig_s_process_elements}.
 
Additionally, we monitored \object{DS\,Aqr} with the CORALIE
spectrograph mounted on the Swiss telescope at La Silla,
Chile. CORALIE \citep{Queloz_1999} is a high resolution (R
$\sim$\,$50000$) fibre-fed echelle spectrograph specifically
constructed to obtain high precision radial velocities through a
cross-correlation technique. We started the monitoring in August 2000
and obtained 40 data points.
 
\subsection{Photometric} 
Using the light curve ephemeredes of \cite{Kholopov_1998}, we estimate
that the spectra of \object{V453\,Oph} and \object{DS\,Aqr} were taken
at pulsation phase $\phi = 0.4$ and $\phi = 0.0$
respectively. Unfortunately, no simultaneous photometric data over a
large wavelength range could be obtained for these phases. Instead we
used photometric data obtained around the phases $\phi = 0.55$ and
$\phi = 0.3$ respectively.  These data are reported in Table
\ref{tab_photometric_data}.
 
The total reddening was determined minimising the difference between
model atmospheres and dereddened fluxes in the optical wavelength
range. For the model atmospheres, Kurucz models with parameters
determined in Sect. \ref{parameter_determination} were used.  This is
possible because small photometric phase discrepancies are only expected 
to result in small shifts in atmospheric parameters.
 
\begin{figure}[h] 
\includegraphics[height=\hsize, angle=-90]{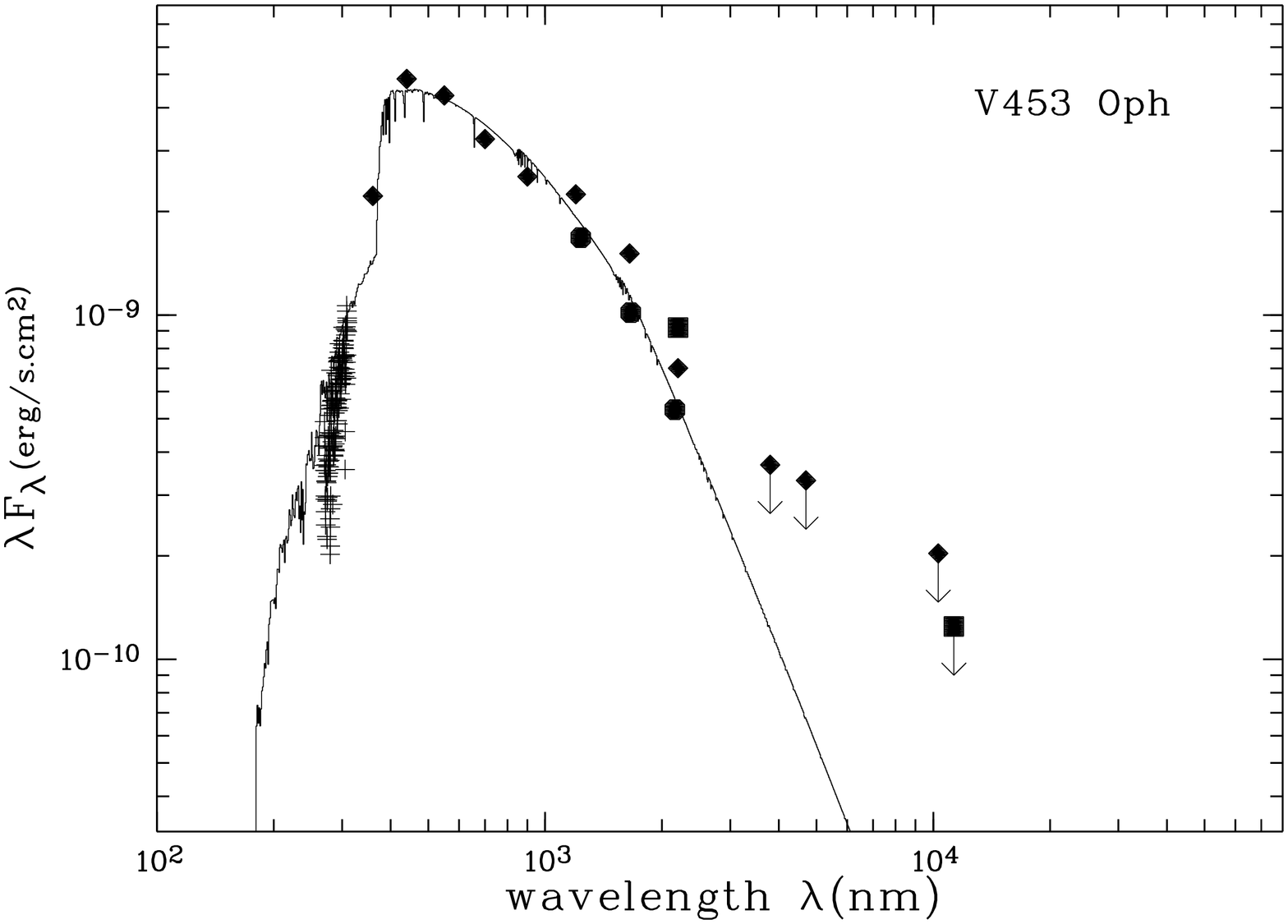} 
\includegraphics[height=\hsize, angle=-90]{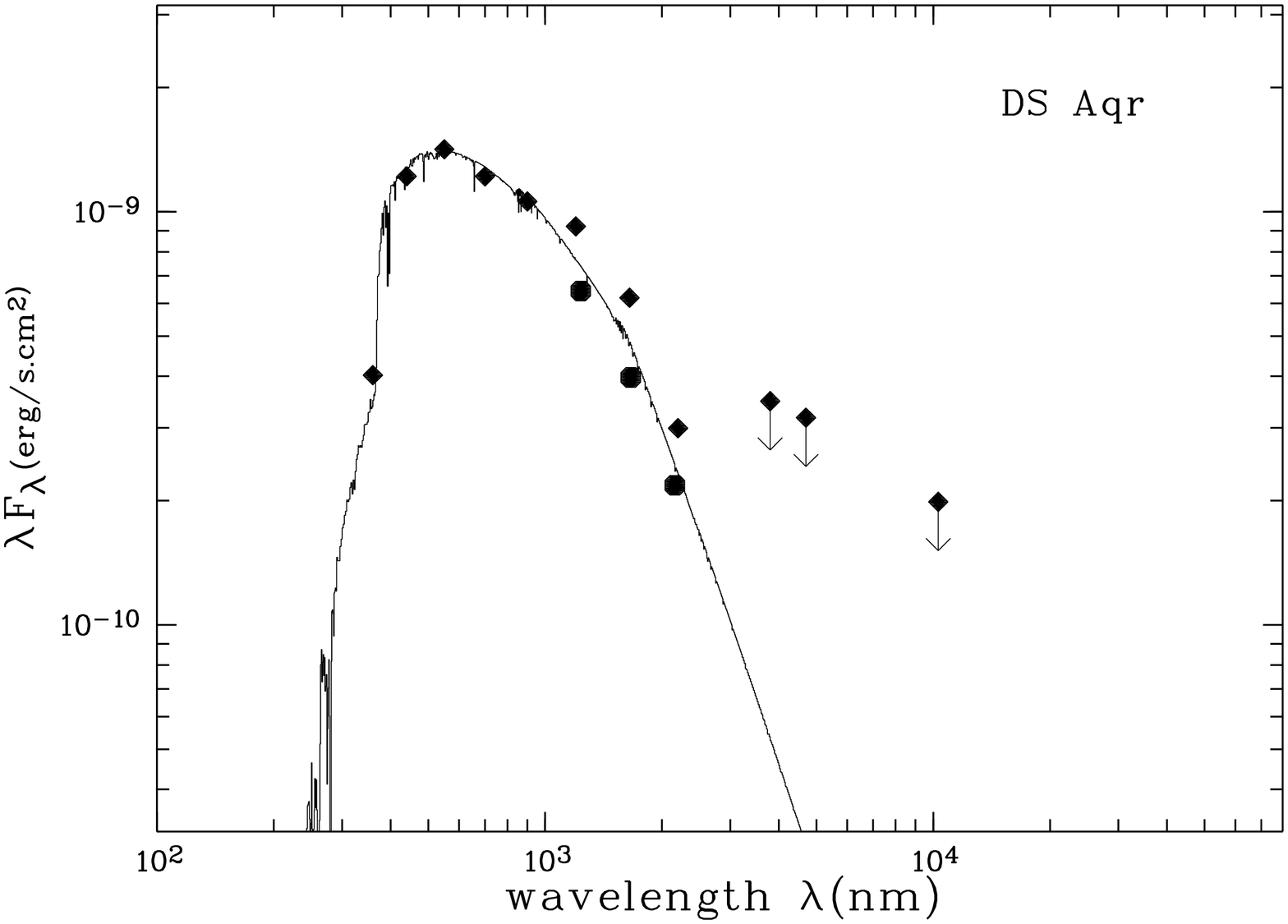} 
\caption{The SED for \object{V453\,Oph} (top) and \object{DS\,Aqr}
(bottom), constructed using a total reddening of E(B-V)$=0.50$, 0.05
respectively. The full line is the Kurucz model, while the other
symbols are photometric measurements from various authors, listed in
Table \ref{tab_photometric_data}. Upper limits are marked with arrows
down. The plus signs are taken from the IUE archive, the diamonds from
\cite{Goldsmith_1987}, the circles from the 2MASS archive and the
squares are from \cite{Gehrz_1972}. The figure shows clearly that no
near-IR excess is observed. }\label{SED_both}
\end{figure}

For \object{V453\,Oph}, we obtain a total reddening of E(B-V)$=
0.50$ and for \object{DS\,Aqr}, E(B-V)$=0.05$. The resulting SED for
both stars is shown in Fig. \ref{SED_both}.  No near-infrared excess
was detected for either variable. The high reddening of
\object{V453\,Oph} is consistent with the E(B-V) obtained from the
diffuse interstellar bands at $\lambda$=5780 \AA\, and 5797 \AA, of
which the equivalent width correlates well with this parameter
\citep{Josafatsson_1987}. Using the data reported in
\cite{Krelowski_1999} these bands give an E(B-V)=0.48 and 0.54
respectively for V453\,Oph.
 
\begin{table}[htp] 
\centering 
\caption{The photometric measurements of \object{V453\,Oph} and
\object{DS\,Aqr}. The data are taken from various authors and plotted
in Fig. \ref{SED_both}: $^{\rm{a}}$ Data from \cite{Goldsmith_1987};
$^{\rm{b}}$ Data from the 2MASS catalogue; $^{\rm{c}}$ Data from
\cite{Gehrz_1972}.}\label{tab_photometric_data}
\begin{tabular}{l|rr|rr} 
\multicolumn{1}{l}{band} & \multicolumn{2}{c}{\object{V453\,Oph}
(mag)} & \multicolumn{2}{c}{\object{DS\,Aqr} (mag)}\\ \hline \hline U
& 11.76$^{\rm{a}}$ & & 11.66$^{\rm{a}}$ &\\ B & 11.33$^{\rm{a}}$ & &
11.18$^{\rm{a}}$ &\\ V & 10.54$^{\rm{a}}$ & & 10.51$^{\rm{a}}$ &\\ R &
9.99$^{\rm{a}}$ & & 10.12$^{\rm{a}}$ &\\ I & 9.39$^{\rm{a}}$ & &
9.73$^{\rm{a}}$ &\\ J & 8.58$^{\rm{a}}$ & 8.806$^{\rm{b}}$ &
9.16$^{\rm{a}}$ & 9.488$^{\rm{b}}$\\ 2.2\,$\mu m$ & & 7.6$^{\rm{c}}$ &
& \\ H & 8.09$^{\rm{a}}$ & 8.442$^{\rm{b}}$ & 8.79$^{\rm{a}}$ &
9.197$^{\rm{b}}$\\ K & 7.95$^{\rm{a}}$ & 8.283$^{\rm{b}}$ &
8.72$^{\rm{a}}$ & 9.089$^{\rm{b}}$\\ L & $>7.0$$^{\rm{a}}$ & &
$>7.0$$^{\rm{a}}$ &\\ M & $>6.3$$^{\rm{a}}$ & & $>6.3$$^{\rm{a}}$ &\\
N & $>4.5$$^{\rm{a}}$ & & $>4.5$$^{\rm{a}}$ &\\ 11.3\,$\mu m$ & &
$>4.6$$^{\rm{c}}$ & &\\
\end{tabular} 
 
\end{table} 
 
\section{Parameter determination}\label{parameter_determination} 
 
\subsection{Atomic data} 
For the abundance analysis, all lines showing a clear symmetric
profile were measured. These lines were identified, based on
wavelength, using mainly the solar line list of
\cite{Thevenin_1989}. This list was also employed to identify blended
lines using the calculated equivalent width (EW) of all lines in the
spectral region. To calculate the abundances of the different
elements, atomic data were preferably taken from the list with
accurate oscillator strengths for A to F type stars, which has been
collected at the Instituut voor Sterrenkunde \citep{Vanwinckel_2000, Reyniers_2004}. Alternatively, to obtain accurate oscillator
strengths, the line lists of the Vienna Atomic Line Database (VALD2)
were used (http://www.astro.univie.ac.at/$\sim$vald/,
\cite{Kupka_1999}). Additionally, the line list of the
D.R.E.A.M. project (http://www.umh.ac.be/$\sim$astro/dream.html) was
 used for Ce, while for La, Eu and Dy, data reported in
respectively \cite{Lawler_2001a}, \cite{Lawler_2001b} and
\cite{Biemont_1993} were used.

\subsection{Radial velocities} 
The heliocentric radial velocities for \object{V453\,Oph} and
\object{DS\,Aqr} were determined using Fe lines as
$-123.8\pm0.6\,\hbox{km$\;$s$^{-1}$}$ and
$-23.9\pm0.8\,\hbox{km$\;$s$^{-1}$}$ respectively. These values differ
respectively only by 5\,\hbox{km$\;$s$^{-1}$}\ and
3\,\hbox{km$\;$s$^{-1}$}\ from published values \citep{Giridhar_1998}.
 
\begin{figure} 
\includegraphics[height=\hsize,angle=-90]{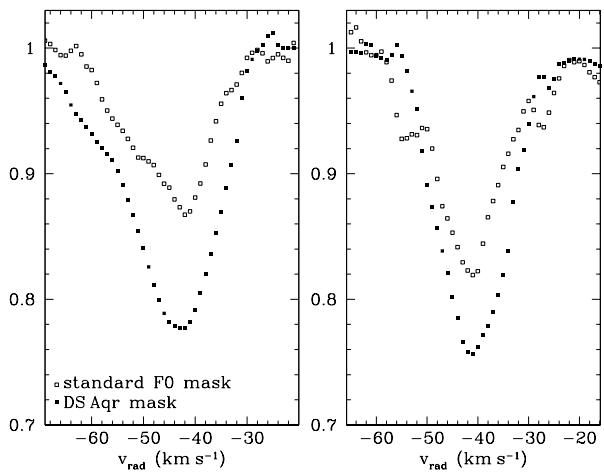} 
\caption{Cross-correlation profiles for \object{DS\,Aqr}. Comparison
of the cross-correlation profiles for two CORALIE spectra, obtained
with a standard mask (open squares) and a specific template for DS Aqr
(full squares). The profiles on the {\em left panel} are for the
spectrum taken on 21/07/2001, the ones on the {\em right panel} for
the spectrum of 25/07/2001.}\label{fig:crossprfls}
\end{figure}

For \object{DS\,Aqr}, we have also 40 high resolution, low
signal-to-noise CORALIE spectra (see
Sect. \ref{observations_spectroscopic}). In the online reduction of
these spectra, radial velocities are determined by cross-correlation
with a standard F0 mask. Since many of the correlation profiles were
quite noisy, we decided to construct a mask optimized for
\object{DS\,Aqr}. This technique is proven to be very efficient for
other RV\,Tauri stars \citep{Maas_2003}. The special mask contains
only the lines that were used in the abundance study
(Sect. \ref{sect_abundance}). In Fig. \ref{fig:crossprfls} we compare
the correlation profiles for two spectra obtained by using the
standard mask and by using our constructed mask. Their smoother
profile and their deeper central depth clearly indicate the higher
quality of the latter. Six out of the 40 spectra were of a
too low signal-to-noise to obtain a reliable velocity.
 
\begin{figure} 
\includegraphics[height=\hsize,angle=-90]{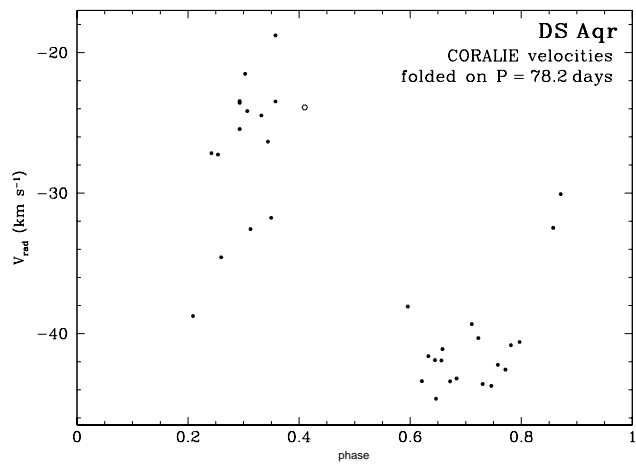} 
\caption{Phase diagram of the 34 CORALIE velocities (filled circles)
and the velocity obtained from the FEROS spectrum (open circle) of
\object{DS\,Aqr}. The velocities are folded on the photometric period
of 78.213 days that was published by \citet{Kholopov_1998}. Since
cycle-to-cycle intrinsic variability is often observed in RV\,Tauri
stars, a much better sampling is needed to be able to test for binary
motion.  }\label{fig:phasedsqr}
\end{figure}

The remaining 34 velocities for \object{DS\,Aqr} display a large
peak-to-peak amplitude of $26\,\hbox{km$\;$s$^{-1}$}$ (see
Fig. \ref{fig:phasedsqr}) Such a high value is not uncommon even for
single RV\,Tau stars, and does not necessarily suggest a
binary nature of the object. The photometric phase distribution of our
34 velocities is poor and since cycle-to-cycle variability is often
observed in RV\,Tauri stars, we have as yet no significant test to
probe the eventual binary nature of \object{DS\,Aqr}.
 
Concerning the radial velocities of the two programme stars, we can
conclude that the variability of the radial velocities can be
accounted for by the known photometric variability of the stars, with
peak-to-peak amplitudes in V of 1.1\,mag for \object{V453\,Oph} and
0.8\,mag for \object{DS\,Aqr}.
 
\subsection{Determination of atmospheric parameters}\label{sect_atmospheric_parameters} 
The abundance pattern was computed using the latest ATLAS9 LTE model
atmospheres \citep{Castelli_2004} in combination with the LTE line
analysis program MOOG of Sneden (version 2002). The model
atmospheres are uniquely determined by the effective temperature
($\hbox{${\rm{T}}_{\rm eff}$}$), gravity ($\log(g)$) and metallicity
([Fe/H]). They are computed with a constant microturbulent velocity
($\xi_t$ = $2\,\hbox{km$\;$s$^{-1}$}$) for the opacity distribution
functions.
 
\begin{figure}[h] 
\includegraphics[height=\hsize, angle=-90]{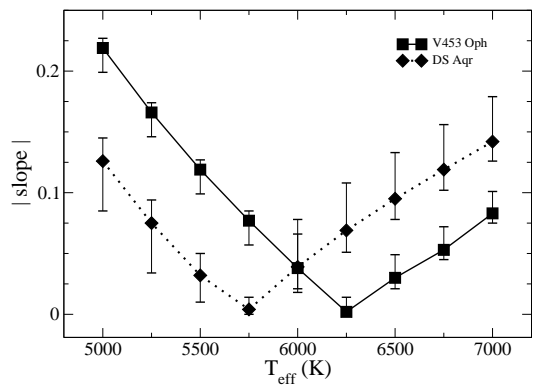} 
\caption{The slope of the trend between the abundance derived from
Fe\,I lines and the EP versus temperature. The values for
\object{V453\,Oph} are depicted using squares and those for
\object{DS\,Aqr} using diamonds. These values were determined using
$\log(g)=1.5$, and $\log(g)=0.5$ respectively. An estimate on the
errors was made by varying the microturbulent velocity by
$\pm1\,\hbox{km$\;$s$^{-1}$}$.}
\label{zero_slope_determination} 
\end{figure}

We derived the photospheric model parameters solely using
spectroscopic criteria. A temperature is determined by forcing the
abundance derived from Fe\,I lines to be independent of their
excitation potential (EP). The large range in EP and the very large
number of lines (\hbox{\object{V453\,Oph}:} $\textrm{EP}=0.0 - 4.4
\,\textrm{eV}$, N$_{\rm{Fe\,I}}=104$, N$_{\rm{Fe\,II}}=32$;
\hbox{\object{DS\,Aqr}:} $\textrm{EP} = 0.0-5.0 \,\textrm{eV}$,
N$_{\rm{Fe\,I}}=267$, N$_{\rm{Fe\,II}}=49$) ensures an accurate
identification of this temperature. This is shown in
Fig. \ref{zero_slope_determination} and confirmed by other elements
with a large number of detected lines (e.g. Ti and Cr).
 
The surface gravity was determined demanding ionisation balance
between Fe\,I\ and Fe\,II\ lines. The microturbulent velocity
was estimated requiring the independence of the abundance derived from
Fe\,I\ lines of the reduced equivalent width. The resulting stellar
parameters are shown in Table \ref{tab_stellar_parameters}, where the
Fe abundance is determined using an input model with a metallicity of
$\textrm{[M/H]} = -2.0$ and $\textrm{[M/H]} = -1.5$ for
\object{V453\,Oph} and \object{DS\,Aqr} respectively.  We used the
conservative errors on the model atmosphere parameters of
$\Delta\hbox{${\rm{T}}_{\rm eff}$}=250 \,\textrm{K}$ and $\Delta
\log(g)=0.5$ (see Fig. \ref{zero_slope_determination}).  The error on
the microturbulent velocity is adopted to be $\Delta \xi_t = 0.5
\,\hbox{km$\;$s$^{-1}$}$.
 
\begin{table}[h] 
\centering 
\caption{The stellar parameters of the stars, as determined using
only spectroscopic criteria. The reported error on the Fe abundance
is only due to line-to-line scatter. See text for more
details.}\label{tab_stellar_parameters}
 
\begin{tabular}[h]{lccccc} 
\multicolumn{1}{l}{Name} & \multicolumn{1}{c}{$\hbox{${\rm{T}}_{\rm
eff}$}$\,(K)}&\multicolumn{1}{c}{$\log(g)$}&
\multicolumn{1}{c}{$\xi_t$\,(\hbox{km$\;$s$^{-1}$})}&\multicolumn{1}{c}{[Fe/H]}\\
\hline \hline \object{V453\,Oph} & 6250 & 1.5 & 3.0 & $-2.23\pm0.12$
\\ \object{DS\,Aqr} & 5750 & 0.5 & 3.5 & $-1.62\pm0.12$
\end{tabular} 
\end{table}

\section{Abundance analysis}\label{sect_abundance} 
 
\subsection{Homogeneous analysis} 
 
Severe selection criteria were used to incorporate or discard an
absorption line in the analysis.  The selected absorption lines have a
small equivalent width (EW$<150\,$m\AA) and a clear symmetric
profile. Blended lines were only taken into account if clearly present
and unblended in the other star. The abundances derived from these
lines are shown in Table \ref{abundance_total}.
 
\begin{figure}
\includegraphics[height=\hsize,angle=-90]{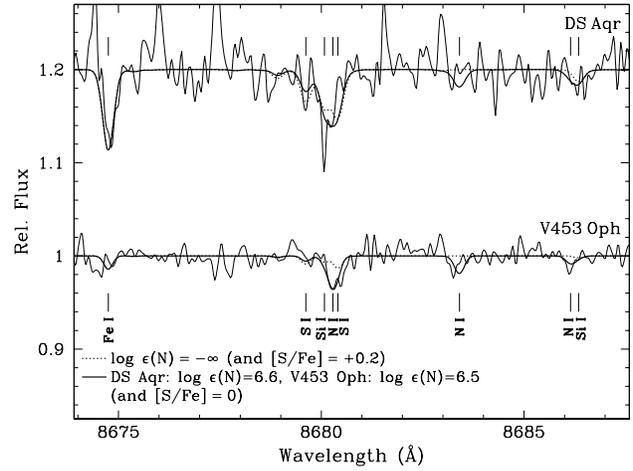}
\caption{Derivation of the upper limit for the nitrogen abundance,
using spectrum synthesis of the N\,I triplet at 8680\,\AA. Two
syntheses for each star were made: one without any nitrogen, to detect
possible blends (dotted line), and the other one with the proposed
upper limit abundance for nitrogen (thick continuous line). From the
first synthesis, it turned out that the strongest N\,I line at
8680.28\,\AA\ is not reliable to deduce an N abundance, due to weak
Si\,I and S\,I blends. Therefore, the upper limits obtained in the
second synthesis focus on the weaker line at 8683.40\,\AA, being
$\log$\,$\epsilon$(N)\,$<$\,6.6 for \object{DS\,Aqr} and
$\log$\,$\epsilon$(N)\,$<$\,6.5 for \object{V453\,Oph}. Log({\it gf}) values
for the three N\,I lines were taken from \citet{Hibbert_1991}.}
\label{fig:nsynth}
\end{figure}

\begin{figure*}[t!] 
\includegraphics[height=\hsize,angle=-90]{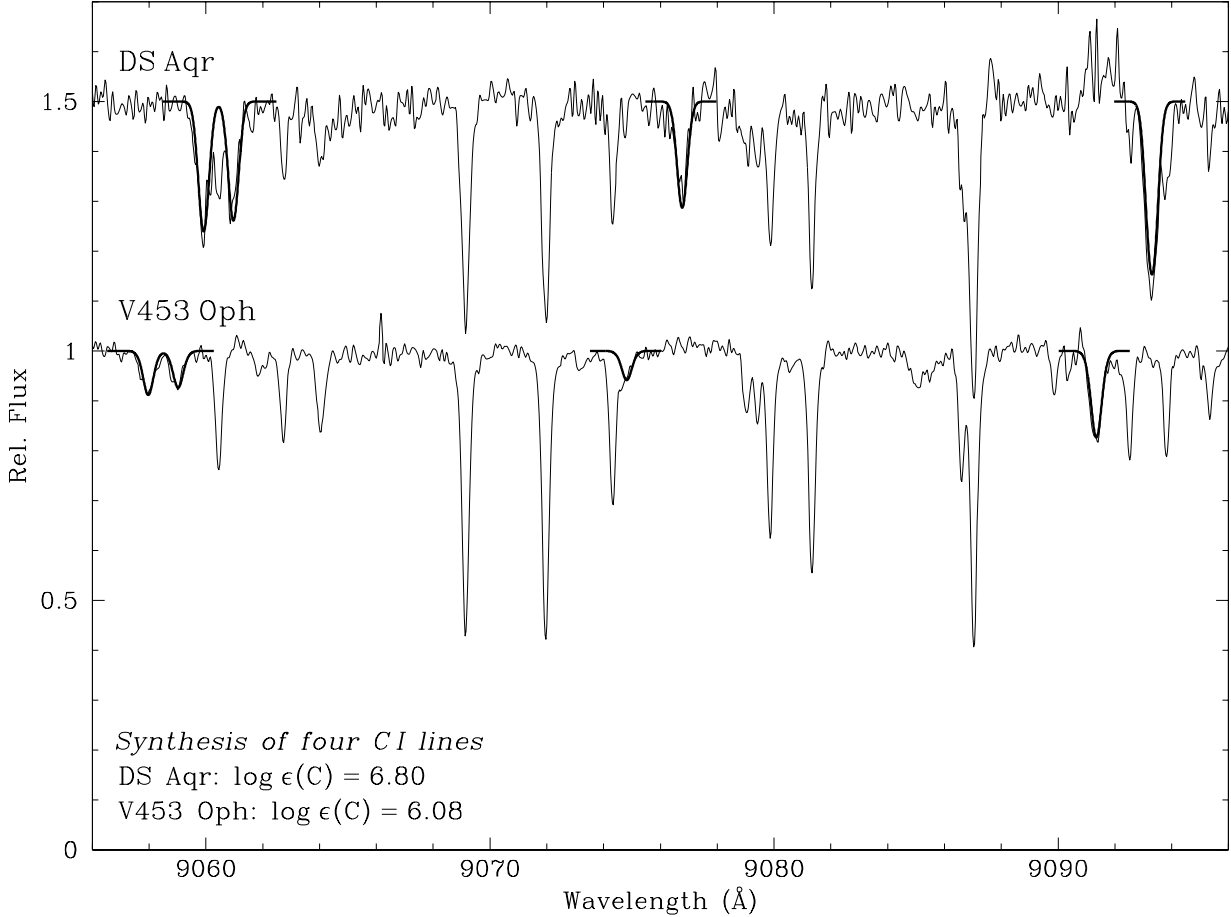} 
\caption{Synthesis of the four C\,I lines used in the abundance
analysis. These four lines turned out to be the only ones from which a
reliable carbon abundance could be derived. The observed spectra are
shown with a thin line; the syntheses with a thick line. The spectra
are not velocity corrected, providing that the telluric lines are
shown at their rest wavelength, and that they can be easily
recognised. The macroturbulent velocity used in the syntheses is
8\,\hbox{km$\;$s$^{-1}$}\ for \object{DS\,Aqr} and
10\,\hbox{km$\;$s$^{-1}$}\ for \object{V453\,Oph}. Log({\it gf}) values for
these four lines were taken from \citet{Hibbert_1993}.}
\label{spectrum_synthese} 
\end{figure*}

The abundance of the CNO elements is always difficult to determine,
particularly for such low metallicity stars. For C, the well-known
multiplet at $7115$\,\AA\ is undetectable for both stars. Therefore, a
spectrum synthesis was performed using four C lines around 9070\,\AA\
of which the result is shown in Fig. \ref{spectrum_synthese}. Telluric
blending, important in this part of the spectrum, is easily identified
by comparing the non-velocity-corrected spectra of both stars, as
shown in Fig. \ref{spectrum_synthese}. A reliable nitrogen abundance
is even more difficult to obtain, since the only N line that is
clearly present in our spectrum is weakly blended by
Si\,I and S\,I lines. As a consequence, we were only able to deduce
upper limits for nitrogen, being $\log$\,$\epsilon$(N)\,$<$\,6.6 for
\object{DS\,Aqr} and $\log$\,$\epsilon$(N)\,$<$\,6.5 for
\object{V453\,Oph}. The derivation of these upper limits is shown in
the synthesis in Fig.\ref{fig:nsynth}. The oxygen abundance is
primarily based on the strong (non-LTE sensitive) 7774\,\AA\
multiplet. For both stars this resulted in a slight underabundance of
C and an overabundance of N and O, as reported in Table
\ref{abundance_total}. This N overabundance is probably a consequence of the
first dredge-up for both stars, while the O overabundance is likely of
initial composition \citep[e.g.][]{Boesgaard_1999}.

Abundances are hard to obtain from sodium (Na) to sulphur (S) due to
the lack of lines. However most of the $\alpha$-elements (Mg, Si, Ca
and Ti) were detected, showing a mean of [$\alpha$/Fe]$=0.3$ for both
stars, which is expected for their metallicity
\citep{Lambert_1987,Gratton_1991}.  The Al abundance found for
\object{V453\,Oph} ([Al/Fe]$=-0.59$) differs significantly from this,
but it is in agreement with the trend of the Al abundance seen in
metal-poor stars \citep[e.g.][]{Gratton_1988}. The intermediate mass
elements from V to Zn are easier to determine. These elements follow
the Fe abundance more closely except for Cr and Mn. While for Mn the
underabundance in both objects is expected
\citep[e.g.][]{Gratton_1991}, the rather large underabundance of Cr in
\object{DS\,Aqr} is not.
 
\begin{figure*}[t!] 
\includegraphics[height=\hsize,angle=-90]{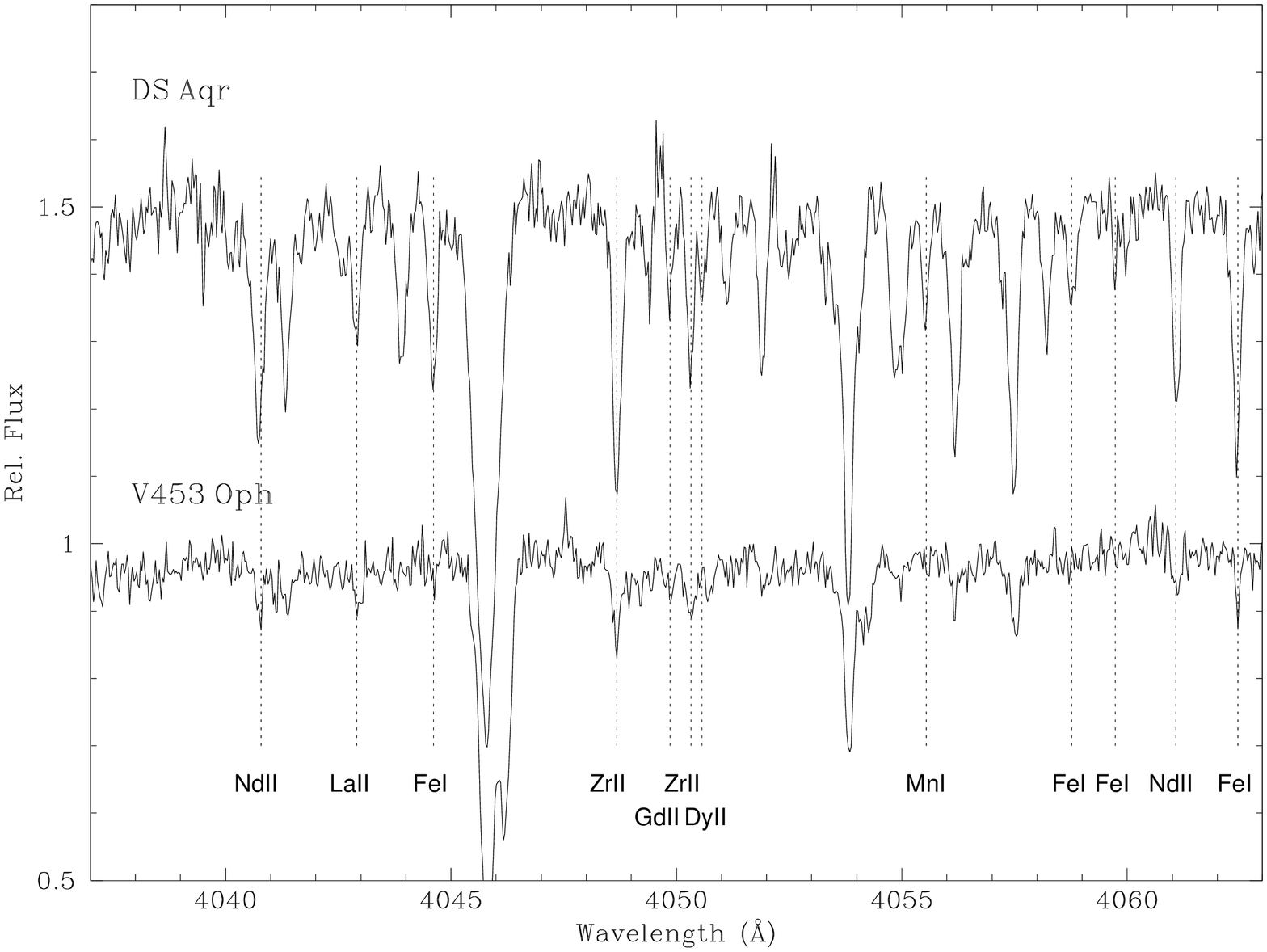} 
\caption{ The velocity corrected spectra of the programme stars
showing some of the used s-process lines. For \object{V453\,Oph}, the
s-process lines are almost the only visible lines in the spectrum.
}\label{fig_s_process_elements}
\end{figure*}

A specific procedure was used to find all available lines for the
s-process elements (Y, Zr, Ba, La, Ce, Nd and Sm).  All lines in the
VALD database were used to estimate their equivalent width for the
assumed stellar parameters. All lines were then measured in the
spectra according to their estimated strength up to the detection
limit (EW$\sim 5$\,m\AA). Using this method, we maximised the number
of lines per species and avoided measuring blends. Some of the
s-process lines used in the analysis are shown in
Fig. \ref{fig_s_process_elements}. Only for Ba was the abundance
estimated using fewer than five lines. The abundance of Ce\,II,
obtained using the atomic data from the D.R.E.A.M. database
\citep{Palmeri_2000} is also reported. Using this data instead of the
VALD data, the abundance of Ce remains fixed for \object{DS\,Aqr},
while it increases for \object{V453\,Oph}.
 
Table \ref{abundance_total} gives the abundances of the
light (ls) and heavy (hs) s-process elements. The light s-elements
have a neutron number close to the magic number 50 (Y and Zr), while
the heavy s-elements cluster around the magic neutron number 82 (Ba,
La, Ce, Pr, Nd and Sm). The indices indicate a mild overabundance of
both ls as hs elements for \object{V453\,Oph}, while for
\object{DS\,Aqr} the s-process elements scale with Fe.

The Eu abundance was determined for both objects. However, Eu lines
are known to have a strong hyperfine splitting (hfs). A reliable Eu
abundance can therefore only be derived when this effect is taken into
account. The Eu abundance of \object{DS\,Aqr} is based on two Eu\,II
lines: the resonance line at 4129.72\,\AA\
(W$_{\lambda}$\,=\,94\,m\AA) and the line at 6645.06\,\AA\ (14\,m\AA);
the Eu abundance of \object{V453\,Oph} is derived solely from the
resonance line (24\,m\AA). A synthesis of these lines including their
hyperfine components showed that the weak lines are only affected by
hfs in their profile, and not in their equivalent width. For the
resonance line in \object{DS\,Aqr}, however, we found a small but
significant effect: the difference in abundance between a hfs and a
non-hfs treatment of this line is $-$0.06\,dex. Hyperfine A and B
constants for this synthesis were taken from \citet{Lawler_2001b} and
the solar isotopic composition of Eu was applied. The Eu abundance
reported in Table \ref{abundance_total} is the hfs-corrected
abundance. A comparison between the Eu and Dy abundance gives [Dy/Eu]
= 0.04 for both stars. This value is in agreement with
\cite{Burris_2000} where a [Dy/Eu] abundance very close to zero was
found for all metallicities (-3.0 $<$ [Fe/H] $<$ -1.0).

\begin{table*}[t!] 
\centering 
\caption{Chemical composition of \object{DS\,Aqr} and
\object{V453\,Oph} as derived from the homogeneous abundance
analysis. For every ion the abundance is listed absolute and relative
to iron. D.R.E.A.M. means that the abundance of Ce\,II is determined
using atomic data obtained from the D.R.E.A.M. database. The reported
Eu abundance is corrected for hyperfine splitting. Also listed is the
solar abundance ($\log( \rm{\epsilon_\odot}) $), taken from the same
references as in \cite{Reyniers_2004} except for Dy, where we used the
solar abundance of \cite{Biemont_1993}. Although more recent data
exist, the values that were used, ensure that the employed $\log({\it gf})$
values are consistent with the solar abundances. The line-to-line
scatter ($\sigma_{\rm{ltl}}$) is reported if more than one line was
used, together with the number of lines used to determine the
abundance (N) and their mean equivalent width
($\overline{\rm{W}}$). Also reported is the error on the abundances
relative to iron ($\sigma_{\rm{tot}}$), determined as discussed in
Sect. \ref{sect_error_analysis}.  This Table ends with the light and
heavy s-process indices (respectively [ls/Fe] and [hs/Fe]) for the
homogeneous analysis. The [ls/Fe] index is the mean of the Y and Zr
abundance and the [hs/Fe] is computed using the mean of the Ba, La and
Nd abundance. Clearly \object{V453\,Oph} shows a significant
overabundance for the s-process elements, while \object{DS\,Aqr} does
not.}
\label{abundance_total} 
 
\begin{tabular}[]{l|cccccc|c|cccccc} 
\multicolumn{1}{c}{}&\multicolumn{6}{c}{}&\multicolumn{1}{c}{}&\multicolumn{6}{c}{}\\
\multicolumn{1}{c}{}&\multicolumn{6}{c}{\large
\textbf{\object{DS\,Aqr}}}&\multicolumn{1}{c}{}&\multicolumn{6}{c}{\large
\textbf{\object{V453\,Oph}}}\\
\multicolumn{1}{c}{}&\multicolumn{6}{c}{$\hbox{${\rm{T}}_{\rm
eff}$}=5750\:K$}&\multicolumn{1}{c}{}&\multicolumn{6}{c}{$\hbox{${\rm{T}}_{\rm
eff}$}=6250\:K$}\\
\multicolumn{1}{c}{}&\multicolumn{6}{c}{$\log(g)=0.5$}&\multicolumn{1}{c}{}&\multicolumn{6}{c}{$\log(g)=1.5$}\\
\multicolumn{1}{c}{}&\multicolumn{6}{c}{$\xi_t=3.5\;\hbox{km$\;$s$^{-1}$}$}&\multicolumn{1}{c}{}&\multicolumn{6}{c}{$\xi_t=3.0\;\hbox{km$\;$s$^{-1}$}$}\\
\multicolumn{1}{c}{}&\multicolumn{6}{c}{[Fe/H]=$-1.62\pm0.12$}&\multicolumn{1}{c}{}&\multicolumn{6}{c}{[Fe/H]=$-2.23\pm0.12$}\\[0.5cm]
 
\hline 

\multicolumn{1}{l|}{ion}& \multicolumn{1}{c}{$\log(\epsilon )$}
&\multicolumn{1}{c}{$\sigma_{\rm{ltl}}$}& \multicolumn{1}{c}{N} &
\multicolumn{1}{c}{$\overline{\textrm{W}}$}&
\multicolumn{1}{c}{[el/Fe]}&\multicolumn{1}{c}{$\sigma_{\rm{tot}}$}&\multicolumn{1}{|c|}{$\log(\epsilon_{\odot})$}&
\multicolumn{1}{c}{$\log(\epsilon )$}
&\multicolumn{1}{c}{$\sigma_{\rm{ltl}}$} & \multicolumn{1}{c}{N} &
\multicolumn{1}{c}{$\overline{\textrm{W}}$}&
\multicolumn{1}{c}{[el/Fe]}&\multicolumn{1}{c}{$\sigma_{\rm{tot}}$}\\
\multicolumn{1}{c|}{}&\multicolumn{3}{c}{}&\multicolumn{1}{c}{m\AA}
&\multicolumn{2}{c|}{}&\multicolumn{1}{c|}{}&
\multicolumn{3}{c}{}&\multicolumn{1}{c}{m\AA}&\multicolumn{2}{c}{}\\

\hline 
\hline 
C\,I	&	6.80		&			&	4		&			&	-0.15		&			&	8.57	 
	&	6.08		&			&       4   		&			&	-0.26		&			\\		 
N\,I	&	$<$6.6		&			&			&			&	$<$+0.2		&			&	7.99	 
	&	$<$6.5		&			&			&			&	$<$+0.7		&			\\ 
O\,I	&	7.88		&	0.17		&	4		&	58		&	+0.64		&	0.16		&	8.86 
	&	7.62		&	0.07		&	3		&	78		&	+0.99		&	0.20		\\ 
Mg\,I	&	6.33		&	0.09		&	5		&	86		&	+0.41		&	0.10		&	7.54 
	&	5.56		&	0.08		&	2		&	51		&	+0.25		&	0.10		\\ 
Al\,I	&			&			&			&			&			&			&	6.47 
	&	3.65		&	0.07		&	2		&	105		&	-0.59		&	0.12		\\ 
Si\,I	&	6.51		&	0.13		&	10		&	16		&	+0.59	&	0.10		&	7.54 
	&			&			&			&			&			&			\\			 
Si\,II	&	6.31		&	0.09		&	2		&	62		&	+0.39	&	0.19		&	7.54 
	&	5.72		&	0.11		&	2		&	23		&	+0.41	&	0.18		\\ 
K\,I	&	3.77		&	-		&	1		&	53		&	+0.27	&	0.16		&	5.12 
	&			&			&			&			&			&			\\ 
Ca\,I	&	4.88		&	0.13		&	24		&	45		&	+0.14	&	0.07		&	6.36 
	&	4.31		&	0.12		&	13		&	21		&	+0.18	&	0.06		\\		 
Sc\,II	&	1.48		&	0.10		&	12		&	64		&	-0.07	&	0.09		&	3.17 
	&	1.22		&	0.10		&	10		&	23		&	+0.28	&	0.06		\\ 
Ti\,I	&	3.55		&	0.08		&	15		&	25		&	+0.15	&	0.06		&	5.02 
	&	3.08		&	0.04		&	4		&	13		&	+0.29	&	0.06		\\ 
Ti\,II	&	3.50		&	0.09		&	37		&	75		&	+0.10	&	0.07		&	5.02 
	&	3.08		&	0.10		&	41		&	27		&	+0.29	&	0.05		\\		 
V\,I	&	2.28		&	0.04		&	2		&	18		&	-0.10	&	0.07		&	4.00 
	&			&			&			&			&		&			\\ 
V\,II	&	2.37		&	0.11		&	7		&	44		&	-0.01	&	0.07		&	4.00 
	&	1.76		&	-		&	1		&	21		&	-0.01	&	0.16		\\ 
Cr\,I	&	3.79		&	0.09		&	8		&	22		&	-0.26	&	0.06		&	5.67 
	&	3.31		&	0.08		&	7		&	44		&	-0.13	&	0.06		\\ 
Cr\,II	&	3.73		&	0.13		&	17		&	39		&	-0.32	&	0.06		&	5.67 
	&	3.35		&	0.13		&	9		&	18		&	-0.09	&	0.06		\\ 
Mn\,I	&	3.35		&	0.16		&	5		&	35		&	-0.42	&	0.09		&	5.39 
	&	2.53		&	0.01		&	2		&	41		&	-0.63	&	0.09		\\ 
Fe\,I	&	5.92		&	0.12		&	267		&	44		&	+0.03	&			&	7.51 
	&	5.30		&	0.11		&	104		&	25		&	+0.01	&			\\ 
Fe\,II	&	5.87		&	0.11		&	49		&	55		&	-0.02	&			&	7.51 
	&	5.27		&	0.13		&	32		&	38		&	-0.02	&			\\ 
Co\,I	&	3.36		&	0.13		&	2		&	56		&	+0.06	&	0.13		&	4.92 
	&	2.88		&	-		&	1		&	19		&	+0.19	&	0.16		\\ 
Ni\,I	&	4.70		&	0.07		&	29		&	20		&	+0.07	&	0.05		&	6.25 
	&	3.89		&	0.05		&	3		&	13		&	-0.13	&	0.06		\\ 
Ni\,II	&	4.68		&	-		&	1		&	25		&	+0.05	&	0.16		&	6.25 
	&			&			&			&			&		&			\\	 
Zn\,I	&	3.11		&	0.06		&	2		&	35		&	+0.13	&	0.06		&	4.60 
	&	2.50		&	-		&	1		&	7		&	+0.13	&	0.16		\\ 
Y\,II	&	0.35		&	0.10		&	13		&	55		&	-0.27	&	0.11		&	2.24 
	&	0.28		&	0.11		&	10		&	26		&	+0.27	&	0.09		\\ 
Zr\,II	&	1.10		&	0.11		&	15		&	50		&	+0.12	&	0.10		&	2.60 
	&	0.98		&	0.12		&	14		&	23		&	+0.61	&	0.09		\\ 
Ba\,II	&	0.42		&	0.05		&	2		&	59		&	-0.09	&	0.16		&	2.13 
	&	0.32		&	-		&	1		&	27		&	+0.41	&	0.22		\\ 
La\,II	&	-0.40		&	0.12		&	9		&	47		&	+0.09	&	0.14		&	1.13 
	&	-0.61		&	0.09		&	6		&	12		&	+0.49	&	0.12		\\ 
Ce\,II	&	-0.12		&	0.09		&	13		&	15		&	-0.08	&	0.14		&	1.58 
	&	-0.19		&	0.07		&	6		&	7		&	+0.46	&	0.12		\\ 
D.R.E.A.M. &	-0.13	&	0.16		&	13		&	15		&	-0.09	&	0.15		&	1.58 
	&	-0.12		&	0.15		&	6		&	7		&	+0.53	&	0.11		\\ 
Nd\,II	&	0.02		&	0.12		&	21		&	24		&	+0.14	&	0.18		&	1.50 
	&	-0.05		&	0.10		&	6		&	13		&	+0.68	&	0.16		\\ 
Sm\,II&	-0.53		&	0.10		&	5		&	13		&	+0.08	&	0.16		&	1.01 
	&			&			&			&			&		&			\\	 
Eu\,II	&	-0.66		&	0.16		&	2		&	54		&	+0.44	&	0.17		&	0.52 
	&	-0.99		&	-		&	1		&	24		&	+0.72	&	0.20		\\ 
Gd\,II	&	-0.10		&	0.11		&	2		&	18		&	+0.40	&	0.14		&	1.12 
	&			&			&			&			&			&			\\
Dy\,II	&	+0.06		&	0.04		&	5		&	24		&	+0.48	&	0.15		&	1.20
	&	-0.35		&	-		&	1		&	17		&	+0.68	&	0.19		\\	 
\hline 
\multicolumn{1}{l}{[ls/Fe]}&\multicolumn{6}{c}{$-0.08\pm0.10$} & \multicolumn{1}{c}{} & \multicolumn{6}{c}{$0.44\pm0.09$}\\ 
\multicolumn{1}{l}{[hs/Fe]}&\multicolumn{6}{c}{$0.05\pm0.15$} & \multicolumn{1}{c}{} & \multicolumn{6}{c}{$0.53\pm0.17$}\\ 
\multicolumn{1}{l}{[hs/ls]}&\multicolumn{6}{c}{$0.13\pm0.18$} & \multicolumn{1}{c}{} & \multicolumn{6}{c}{$0.09\pm0.19$}\\ 
\end{tabular} 
\end{table*}

\subsection{Error analysis}\label{sect_error_analysis}
Considering the absolute abundance determination ($\log(\epsilon)$),
the line-to-line scatter ($\sigma_{\rm{ltl}}$) represents the variance
of the obtained values. Hence, if more than about five lines were
used in the abundance determination, this scatter is
representative of the non-systematic errors on the abundances. As
indicated in Table \ref{abundance_total}, the typical line-to-line
scatter is $0.1$\,dex.

Systematic and consequently more important errors on the derived
abundances are mainly due: uncertainties on the determination of
atmospheric parameters, systematic errors in the $\log({\it gf})$ values and
the presence of non-LTE effects. To determine how sensitive the
abundances are to the stellar parameters, the abundances were
recalculated changing T$_{\rm{eff}}$, $\log(g)$ and
$\zeta_{\rm{t}}$ with the assessed error (see
Sect. \ref{sect_atmospheric_parameters}). From this test, it appears
that the abundance determination is most sensitive to the effective
temperature. The total error due to uncertainties in stellar
parameters is given by adding all independent contributions:
\begin{equation}
\sigma_{\rm{mod}} =
\sqrt{\sigma_{\rm{T}_{\rm{eff}}}^2+\sigma_{{\log(g)}}^2+\sigma_{\zeta_{\rm{t}}}^2}
\end{equation}

The systematic errors due to the possibly biased $\log({\it gf})$ values and
due to non-LTE effects are more difficult to determine. Because of the
similar effective temperature and gravity of both stars, a large
reduction of these errors is guaranteed in a relative abundance
analysis (see Sect. \ref{sect_s-process_overabundances}).

In Table \ref{abundance_total}, the total error on the mean
($\sigma_{\rm{tot}}$) of the [el/Fe] values is shown. This error is
determined by adding the errors on the mean due: uncertainties in
the model atmosphere ($\sigma_{\rm{mod}}$), the line-to-line scatter
on the absolute abundances (defaulted to $\sigma_{\rm{ltl}}$\,=\,0.15
if only one line was used) and the error on the Fe abundance
($\sigma_{\rm{Fe}}$). The employed formula is:
\begin{equation}\label{error_formule}
\sigma_{\rm{tot}} = \sqrt{(\frac{\sigma_{\rm{ltl}}}{\sqrt{\rm{N}_{\rm{el}}}})^2 + (\sigma_{\rm{mod}})^2 + (\frac{\sigma_{\rm{Fe}}}{\sqrt{\rm{N}_{\rm{Fe}}}})^2 }
\end{equation} 

\section{Discussion of chemical analysis}\label{sect_discussion} 
\subsection{Comparison with the literature} 
The chemical patterns of both stars were previously analysed in the
literature. For \object{DS\,Aqr}, two analyses were performed
\citep{Giridhar_1998, Giridhar_2000}, with the latter one being more
elaborate, while for \object{V453\,Oph} only one analysis was
performed \citep{Giridhar_1998}.
 
The analysis in the literature for \object{V453\,Oph} and the one
presented here resulted in the same metallicity and very similar
[el/Fe] values. Only for [C/Fe] and [O/Fe], being respectively
0.5\,dex lower and 0.8\,dex higher in this analysis, there are
discrepancies. Neither previous abundances are compatible with our
spectra.  For the other elements reported in both analyses, the mean
difference relative to Fe is only 0.12\,dex. The high radial velocity,
indicating membership of pop.\,II stars, strengthens the very low
metallicity of the star \object{V453\,Oph}.
 
The metallicity reported in \cite{Giridhar_2000} for \object{DS\,Aqr},
[Fe/H]$=-1.1$, differs significantly from the one reported here.
However our analysis uses a much larger number of lines of
Fe\,I and Fe\,II , so we are confident in the established
value. For the fractional abundances, however, the [el/Fe]
values in \cite{Giridhar_2000} differ on average by only 0.06\,dex from
those reported here. Since we focused on s-process lines, we use many
more available lines to derive quantitative data.
 
We conclude that our spectra are clearly not compatible with some
absolute abundance values found in the literature. Our S/N is
higher and our abundance analysis covers many more elements, in
particular more lines for s-process elements.
 
\subsection{Depletion process} 
\begin{figure}[h] 
\includegraphics[height=\hsize,angle=-90]{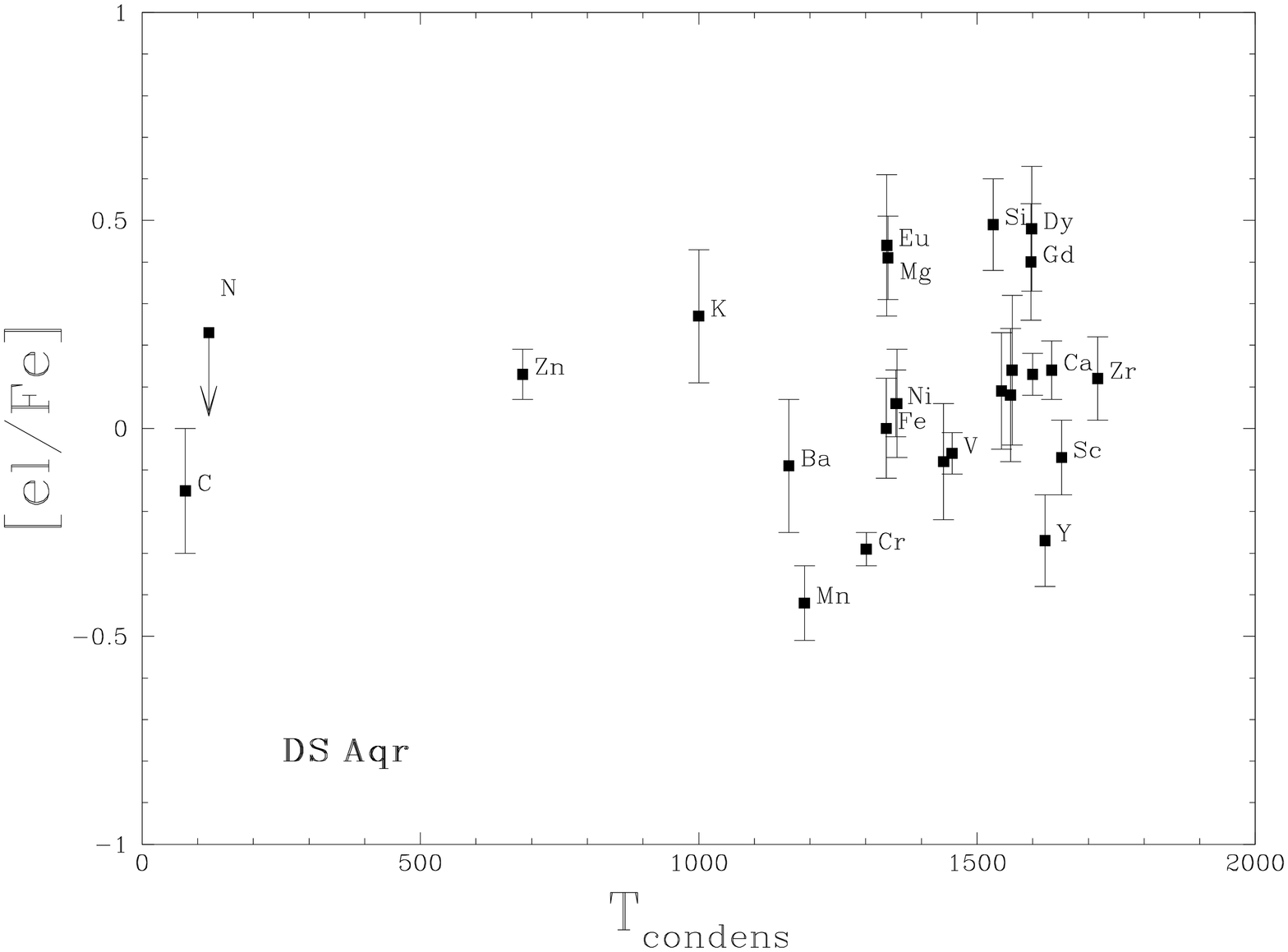}
\includegraphics[height=\hsize,angle=-90]{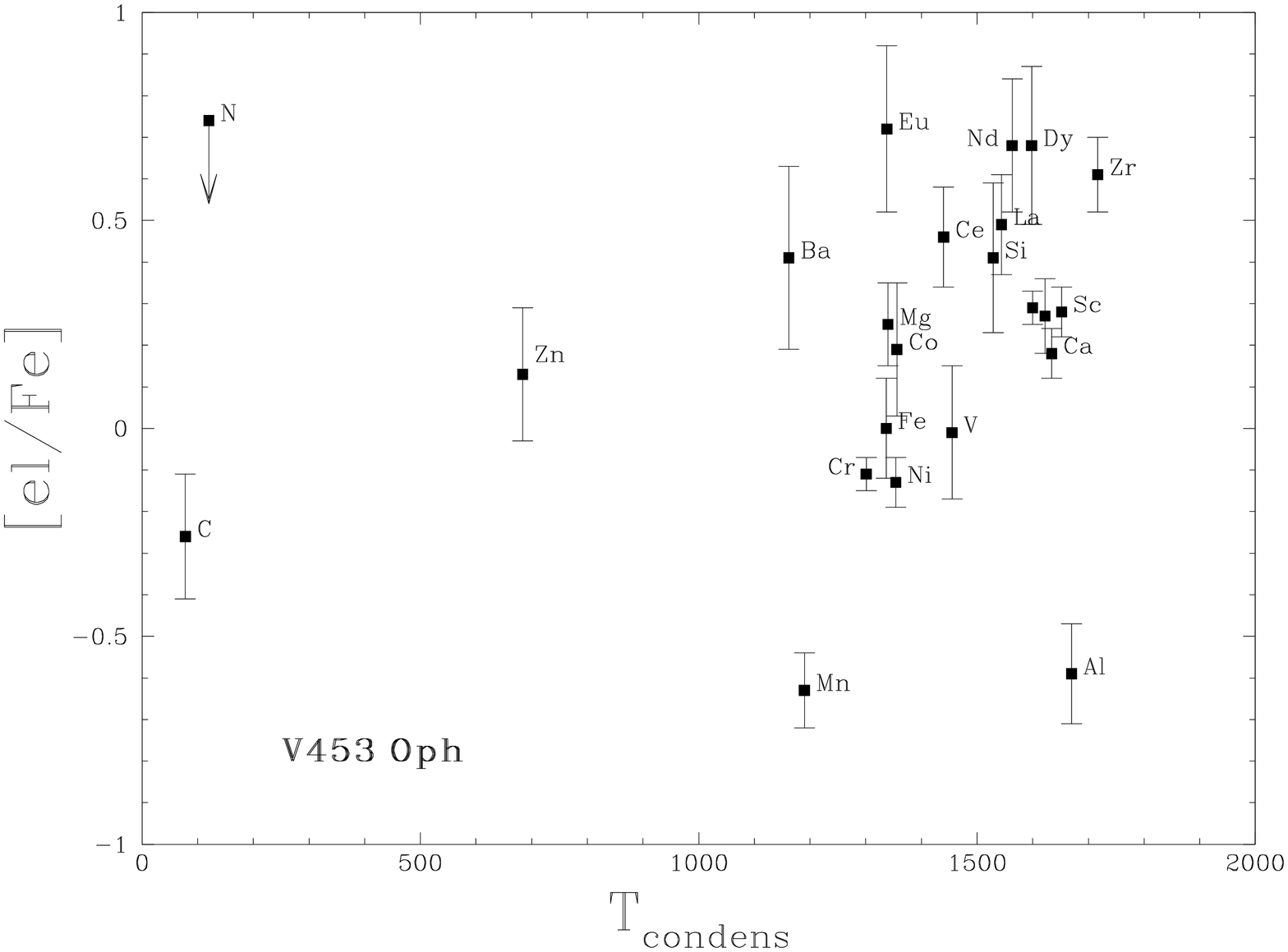}
\caption{The derived abundances versus the condensation
temperature for  \object{DS\,Aqr} (top) and \object{V453\,Oph}
(bottom). These figures indicate that no depletion process
occurred.}\label{fig_tcondens}
\end{figure}

As expected for RVC variables, the depletion process was not efficient
in the investigated objects. This is clearly demonstrated in
Fig. \ref{fig_tcondens}, where the abundances are plotted versus the
condensation temperature \citep{Lodders_1988}. A star subjected to the
depletion process shows a lowered abundance of elements with a
higher condensation temperature. This is clearly not observed in
either variable. Therefore, the derived abundances represent the
initial composition of the stars, possibly altered by their internal
evolution and/or accretion of enriched material from a companion.
 
\subsection{s-process overabundances}\label{sect_s-process_overabundances} 
The s-process indices reported in Table \ref{abundance_total} in
combination with the abundances of the individual s-process elements
indicate that \object{DS\,Aqr} shows the expected abundances for a
photosphere unaltered by internal evolution \citep[e.g.][and
references therein]{Francois_2003}, while for \object{V453\,Oph} the
overabundance of heavy elements indicates mixing of nucleosynthetic
elements in the photosphere.

To constrain possible systematic uncertainties in the reported
abundances, a strictly relative analysis was performed, which uses
only spectral lines detected in both objects. The results of this
analysis are listed in Table \ref{abundance_relative}. By comparing
\object{V453\,Oph} with the results of \object{DS\,Aqr}, we assume
that the latter is a good tracer of the mean Galactic trend.  The
individual relative abundances given in Table \ref{abundance_relative}
($\Delta$[el/Fe]) are thus to be interpreted as the abundance
characteristics of \object{V453\,Oph} with respect to unevolved
objects at similar metallicity that follow the mean Galactic
trend. This analysis provides a solid basis for the proposition that
\object{V453\,Oph} is a metal deficient pop\,II RV\,Tauri star with a
significant enrichment in heavy elements.

The errors on the mean of the relative abundances
($\sigma_{\rm{tot}}$) were estimated assuming that uncertainties due
to the model atmospheres cancel out because both stars have similar
atmospheric parameters. Thus we used a formula similar to
Eq.~\ref{error_formule}, excluding the uncertainty on the model
atmospheres, but including the uncertainty on the Fe abundance of both
stars. The resulting errors on the relative abundances are shown in
Table \ref{abundance_relative}.

\begin{table}[t] 
\centering
\caption{Results from the relative analysis of \object{DS\,Aqr} and
\object{V453\,Oph} only using the lines that were observed in both
spectra. In the column labelled $\Delta$[el/Fe], the difference in
abundance relative to Fe for \object{V453\,Oph} minus the one for
\object{DS\,Aqr} is given. The other notations are similar as in Table
\ref{abundance_total}. This analysis shows clearly an overabundance of
s-process elements in \object{V453\,Oph} because possible bias effects
due to uncertain $\log({\it gf})$ values are eliminated in a relative
study.}
\label{abundance_relative}
\begin{tabular}{l|cccc}
\multicolumn{1}{c}{}&\multicolumn{4}{c}{\large \textbf{Relative}}\\
\multicolumn{1}{c}{}&\multicolumn{4}{c}{\large
\textbf{Analysis}}\\[0.5cm] \hline
\multicolumn{1}{l|}{ion}&\multicolumn{1}{c}{$\sigma_{\rm{ltl}}$}&\multicolumn{1}{c}{$\Delta$[el/Fe]}&\multicolumn{1}{c}{N}&\multicolumn{1}{c}{$\sigma_{\rm{tot}}$}\\
\hline \hline

Mg\,I		&       0.04            &	-0.13		&	2		&	0.03		\\ 
Si\,II		&       0.02            &	+0.02		&	2		&	0.02		\\ 
Ca\,I		&       0.10            &	+0.11		&	12		&	0.03		\\ 
Sc\,II		&       0.10            &	+0.33		&	7		&	0.04		\\ 
Ti\,I		&       0.01            &	+0.20		&	2		&	0.01		\\ 
Ti\,II		&       0.07            &	+0.18		&	24		&	0.02		\\ 
V\,II		&       -               &	-0.02		&	1		&	0.15		\\ 
Cr\,I		&       -               &	+0.25		&	1		&	0.15		\\ 
Cr\,II		&       0.06            &	+0.25		&	8		&	0.02		\\ 
Co\,I		&       -               &	+0.24		&	1		&	0.15		\\ 
Ni\,I		&       0.04            &	-0.16		&	2		&	0.03		\\ 
Zn\,I		&       -               &	+0.04		&	1		&	0.15		\\ 
Y\,II		&       0.12            &	+0.53		&	8		&	0.04		\\ 
Zr\,II		&       0.09            &	+0.47		&	11		&	0.03		\\ 
Ba\,II		&       -               &	+0.54		&	1		&	0.15		\\
La\,II		&       0.10            &	+0.41		&	4		&	0.05		\\
Ce\,II		&       0.02            &	+0.52		&	2		&	0.02		\\ 
D.R.E.A.M.	&       0.02            &	+0.52		&	2		&	0.02		\\ 
Nd\,II		&       0.16            &	+0.55		&	5		&	0.07		\\ 
Eu\,II		&       -               &	+0.39		&	1		&	0.15		\\
\hline
\multicolumn{1}{l}{[ls/Fe]}&\multicolumn{4}{c}{$0.50\pm0.04$}\\ 
\multicolumn{1}{l}{[hs/Fe]}&\multicolumn{4}{c}{$0.50\pm0.10$}\\ 
\multicolumn{1}{l}{[hs/ls]}&\multicolumn{4}{c}{$0.00\pm0.11$}\\ 
\end{tabular}
\end{table}

\subsection{Comparison with CH-giants}

We performed a comparison of the V453 Oph s-enrichment with those
reported for CH stars of similar metallicity. CH-stars are cool C-rich
objects of low metallicity showing enhanced s-process lines. A
distinction is made between early-type and late-type CH-stars:
Early-type CH-stars are probably binaries which are extrinsically
enriched \citep[e.g.][]{Mcclure_1990}, while late-type CH-stars
combine bright absolute magnitudes and high $^{12}$C/$^{13}$C ratios
and are therefore believed to be intrinsically enriched
\citep{Vaneck_2003}.

\begin{figure}[t] 
\includegraphics[height=\hsize, angle=-90]{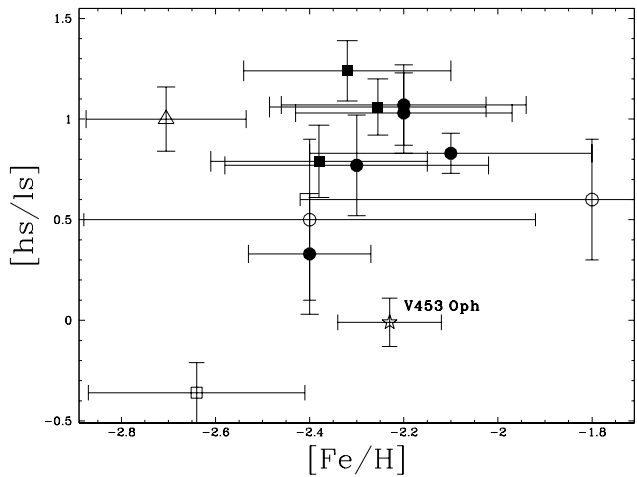} 
\caption{In a narrow metallicity interval, the literature [hs/ls]
values and those for \object{V453\,Oph} are plotted. We refer for the
definition of the ls and hs indexes to the text. The CH stars with the
expected Pb overabundance are depicted with filled symbols, while the
others are represented using open symbols. The circles are taken from
\cite{Vaneck_2003}, the upper triangles from \cite{Aoki_2001} and the
squares from \cite{Aoki_2002}. The star \object{V453\,Oph} is shown as
an asterisk and labelled in the graph. For this star, no Pb abundance
was determined.}\label{fig_ch_giants}
\end{figure}

Standard s-process scenarios predict the production of mainly Pb for
the low-metallicity CH-stars \citep[Pb/Ba ratios of 70 are
predicted for \rm{[Fe/H]}$=-1.3$ by][]{Goriely_2000}. Some of the
studied CH-stars show this extreme Pb enrichment, while others do not
\citep[e.g.][and references therein]{Vaneck_2003}. A comparison of the
reported s-process enrichments is shown in
Fig. \ref{fig_ch_giants}. In this figure, the [hs/ls] value is plotted
as a function of metallicity. The ls value is taken as the abundance of
Zr and the hs value as the mean of the abundances of La, Ce and
Nd. Only stars with all these abundances determined were taken into
account. In Fig.  \ref{fig_ch_giants} a large spread in [hs/ls] is
observed for very similar metallicities. Although V453 Oph is situated
in a less dense region of this plot, some CH-giants have similar
[hs/ls] indices. Extremely Pb-enriched as  well as not Pb-enriched
objects are found with a [hs/ls] value similar to V453 Oph. Note that
the determination of the Pb abundance in the hot V453 Oph star is difficult and
only an upper limit of [Pb/Fe] $<$ 2 could be determined.

\subsection{Other RV\,Tauri stars with s-process enrichment}
To our knowledge, up to now only one other RV\,Tauri object with a
clear s-process enrichment has been detected: V1 in $\omega$\ Cen
\citep{Gonzalez_1994}. This RV Tau star with a spectroscopic class on
the B-C boundary \citep{Wallerstein_1984} also shows mild s-process
overabundances ([ls/Fe]\,=\,0.6 and [hs/Fe]\,=\,0.5). It is unclear
whether this enrichment is intrinsic, or if it is due to the well-studied but poorly understood self-enrichment of $\omega$\ Cen
\citep[e.g.][and references therein]{Tsujimoto_2003}. The fact that
the less evolved variables V48 and V29, also studied in
\citet{Gonzalez_1994}, display the same s-process overabundances as
V1, is in favour of the self-enrichment scenario. V1, however, also
shows a small, but significant, overabundance in carbon compared to
these two other variables ($\Delta$[C/Fe]=0.9 and 0.5 for V48 and V29
respectively). Thus it appears that it is not only possible to enhance
solely the s-process elements (\object{V453\,Oph}) in RV\,Tauri
variables, it is also possible to enhance only the C abundance
($\omega$ Cen V1) without strongly enriching the star with
s-process elements.

\section{V453 Oph, initial, extrinsic or intrinsic heavy element overabundances ?}

The most important abundance characteristic of \object{V453\,Oph} is
that it is a genuine low-metallicity star with [Fe/H]=$-$2.2 showing a
significant heavy element overabundance, but without any C-enrichment.
Moreover, the N-abundance is too low to explain the C deficiency by CN
cycling.  The luminosity and atmospheric parameters of
\object{V453\,Oph} are compatible with an object of low initial mass
(typically $\rm{M}\simeq 0.8$~\msun) in its post-AGB phase of
evolution.

We explored several possible scenarios to explain the
observed abundances: special chemical characteristics of the parental
cloud; enrichment by a binary companion and intrinsic s-process
enrichment by dredge-up.

\subsection{Parental cloud anomalies}

At low metallicity, the Galaxy was not well mixed so there is a large
star-to-star scatter in the initial abundance ratios of unevolved
stars of low iron abundance ([Fe/H] $\leq$ -2). The chemical
composition of low-metallicity stars may reflect strong initial
heavy element enrichment of the parental cloud. To test whether this
may have been the case for \object{V453 Oph}, we scanned the recent
literature for good quality data of individual non-evolved stars. We
compared both the overabundances of individual neutron capture
elements as well as the observed distribution, since the latter
is a good tracer of the nucleosynthesis production site.

\begin{figure}[t]
\includegraphics[width=\hsize]{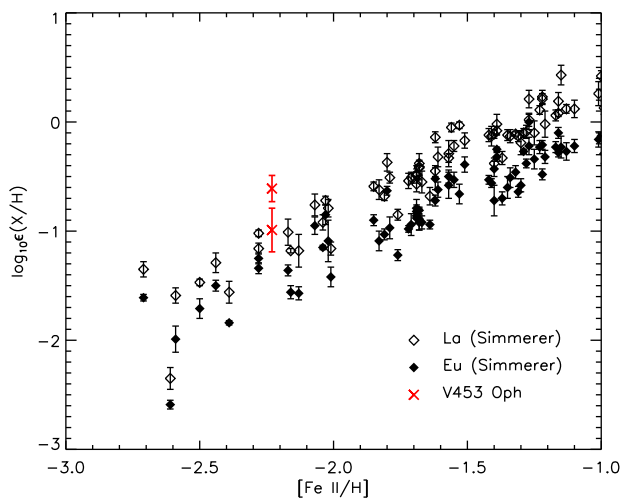}
\caption{A comparison of the La and Eu abundance of V453\,Oph with
  the values found for non-enriched objects in
  \cite{Simmerer_2004}. Although the excess is small,
  the La abundance of V453\,Oph is above the values observed
  for non-enriched objects.}
\label{fig:simmererSample}
\end{figure}

Massive stars are believed to be major contributors of neutron capture
elements in the Galaxy, with a most probable enrichment in r-process
nuclei produced during the supernova explosion, but also of the
so-called weak s-process component responsible for the production of
s-nuclei up to $A\simeq 90$ during core He-burning. Low-metallicity
stars polluted by supernova-enriched material are therefore likely to
show, in addition to an s-process enrichment in elements up to Zr, a
strong r-process signature.  In recent years, it became clear that
this picture is oversimplified and recent observations of a large
sample of low-metallicity stars show evidence of the main and strong
components of the s-process in individual unevolved stars with
metallicities as low as [Fe/H]=$-$3 \citep{Simmerer_2004}. These
authors focus on the study of La and Eu abundances of about 160 giant
and dwarf stars in the metallicity range $-$3$<$[Fe/H]$<$+0.3. In
Fig.\ref{fig:simmererSample} we compare the abundances of
\object{V453\,Oph} with the sample of \citet{Simmerer_2004}.  The La
enrichment is above the level observed in objects of similar
metallicity, but the excess is mild.

Detailed chemical studies of very low-metallicity unevolved stars
tend to focus on the few objects which are known to show this clear
dominant r-process signature. The most famous example in this respect
is the ultra-metal-deficient star CS 22892-052 \citep[][and references
therein]{Sneden_2003}.  We compared the heavy element
distribution of \object{V453\,Oph} with the well studied r-process
enriched halo giant BD$+$17$^{o}$.3248 \citep{Cowan_2002}. Both stars
have the same metallicity and show similar overabundances of some
individual heavy elements.  The difference between both distributions
is most apparent in Eu and Dy. When both distributions are normalised
to Zr, both the Eu and Dy abundances of BD$+$17$^{o}$.3248 are 0.6 dex
larger than what is observed in \object{V453\,Oph}. 

In Table~\ref{comparison} we compare the \object{V453\,Oph} overabundances
to those found in non-evolved objects of similar metallicities from the
extensive study of \citet{Burris_2000}.  At metallicity [Fe/H]=$-$2.2, the
Y abundance of V453 Oph (
[Y/Fe]=$+$0.3 and [Y/Fe]=+0.5 in \object{DS\,Aqr}) is clearly
above the mean value of $-0.2$ from \citet{Burris_2000} sample.  Outliers
to this mean Galactic 
trend exist, however, but the highest observed value in this sample
of 10 stars is about [Y/Fe]=+0.0 in the metallicity range of
\object{V453\,Oph}. The scatter in the [Zr/Fe] ratio at metallicity
around $-$2.2 is similar, but here again  \object{V453 Oph} is
significantly more Zr-rich than the other objects. The La overabundance with respect to the
Galactic mean is confirmed in this comparison. For the heavy elements Nd,
Eu and Dy, the general picture is not as clear. In \object{V453\, Oph},
they tend to be as abundant as in the  \citet{Burris_2000} sample,
although some stars do exhibit similar enrichments.

\begin{table}[t] 
\centering
\caption{Comparison between the enrichments found in
\object{V453\,Oph} and the abundances found in unevolved low-metallicity
stars compiled by  \citet{Burris_2000}.  The mean and standard deviation of
the values 
obtained for the 18 stars in the metallicity range $-2.4\, <\, $ 
[Fe/H]$\, <\, -2.0$ is
given. Not all data are available for all stars, so the number of stars
used in the compilation is given in the last column.}
\label{comparison}
\begin{tabular}{l|c|ccc} \hline
        & \object{V453\,Oph}  &  mean  & 1 $\sigma$ & n \\ \hline \hline
\multicolumn{1}{l|}{[Y/Fe]}  & +0.27  &  $-$0.23 & 0.18 & 10 \\
\multicolumn{1}{l|}{[Zr/Fe]} & +0.61  &  $+$0.17 & 0.25 & 10 \\
\multicolumn{1}{l|}{[Ba/Fe]} & +0.41  &  $+$0.12 & 0.50 & 18 \\
\multicolumn{1}{l|}{[La/Fe]} & +0.49  &  $+$0.05 & 0.28 & 10 \\
\multicolumn{1}{l|}{[Nd/Fe]} & +0.68  &  $+$0.14 & 0.38 & 10 \\
\multicolumn{1}{l|}{[Eu/Fe]} & +0.72  &  $+$0.27 & 0.38 & 12 \\ 
\multicolumn{1}{l|}{[Dy/Eu]} & +0.68  &  $+$0.37 & 0.35 & 10 \\
\end{tabular}
\end{table}

In conclusion, although the [Eu/Fe] and [Dy/Eu] abundances in both
\object{V453\,Oph} and \object{DS\,Aqr} indicate an initial enrichment of
r-process nature, the relatively large Y, Zr, Ba and La overabundances of
\object{V453\,Oph} with respect to \object{DS\,Aqr} and the low-metallicity
stars in the \citet{Burris_2000} sample  clearly points to an additional
s-process enrichment in \object{V453\,Oph}. If the actual chemical
distribution of \object{V453 Oph} reflects the initial composition of the
parental cloud, this cloud must have been subject to some s-process
enrichment capable of producing elements up to the Ba peak. However, up to
now, only AGB stars are believed to be responsible for this corresponding
main component of the s-process, massive stars being capable of producing
elements up to Zr only. 

\subsection{Extrinsic enrichment}

The enrichment we see in \object{V453 Oph} may come from the  pollution of
a putative companion similar to the extrinsically enriched CH
stars discussed in Section 5.4.  Binarity among dusty RV\,Tauri stars
may be widespread \citep{vanwinckel_1999, Deruyter_2004} and the
arguments are either direct (with direct radial velocity measurements), or
indirect (SED characteristics pointing to the presence of a dusty disc
instead of an outflow). Note that only the dusty RV\,Tauri stars are
thought to be binaries in which the companion is probably not a white
dwarf but an unevolved main sequence star.  No binary RV\,Tauri star
is definitely identified with a White Dwarf companion.

Unfortunately, we do not have detailed radial velocity monitoring data
for V453 Oph.  In V453 Oph, no dust-excess is observed so the
above mentioned indirect evidence for binarity is lacking.  Also, the
intrinsic luminosity determined by the P-L relation is large, pointing
to a genuine post-AGB nature. Note that extrinsically enriched objects are rare : Giant
barium stars (extrinsically enriched Giants with white dwarf
companions) represent only 1 \% of the giant G-K stars
\citep[e.g.][]{Jorissen_2004}.

We conclude that, although we cannot exclude the possibility that
\object{V453 Oph} belongs to a binary system which has been polluted by an
AGB star that is now a White Dwarf companion, this hypothesis is not very
likely. Moreover, the C deficiency associated with the s-process enrichment
remains difficult to explain by standard nucleosynthesis models (see
Sect. 7). This scenario presents however the advantage that the White Dwarf
progenitor responsible for the s-process nucleosynthesis can a priori be of
any initial mass (below some 8~\msun). 

\subsection{Intrinsic enrichment}

The most natural scenario for a post-AGB star to become overabundant
in s-process elements is that the object was subject to internal
nucleosynthesis followed by dredge-up episodes during its previous AGB
phase. The most favoured s-process model is associated with the
partial mixing of protons (PMP) into the radiative C-rich layers at
the time of the 3DUP \citep{Gallino_1998, Goriely_2000}. There is,
however, no general agreement on what physical mechanism can be
responsible for the injection of protons into the \chem{12}C-rich
layers. Furthermore, in this scenario, the s-process surface
enrichment is bound to be associated with a stronger C pollution.

With a metallicity of $-$2.2 and a luminosity $\rm{L}=2400\pm
1600$~\lsun, this post-AGB star must have evolved from a low-mass main
sequence star (typically $\rm{M}\simeq 0.8$~\msun).  For such stars, it
remains unknown if any PMP may take place in the C-rich region, and
even if any third dredge-up can intrinsically enrich the stellar
surface. The results of our abundance analysis are consequently
difficult to reconcile with this evolutionary path as well.


\section{S-process models and observations}

As explained in the previous section, none of the standard scenarios
can a priori explain the present observations. It is clear that the
abundances of \object{V453 Oph} challenge our current understanding of
the s-process. On the one hand, s-process nucleosynthesis during core
He-burning in massive stars can not account for a significant
production of s-elements up to the Ba peak. On the other hand, any AGB
s-process models within the PMP scenario would lead to a concomitant
and significant production of C which is not observed. More
specifically, as described in detail in \citep{Deroo2005} , the PMP
model in the intrinsic enrichment scenario could in principle lead to
an abundance distribution close to the observed one, but would
simultaneously pollute the stellar envelope with a C overabundance of
the order of [C/Fe]=1.5. The major specificity of this PMP s-process
is the prediction of a large Pb overabundance that cannot be confirmed
by the present observations (the upper limit established from the
observations is [Pb/Fe]$\leq2.0$).

\begin{figure}
\includegraphics[height=\hsize,angle=-90]{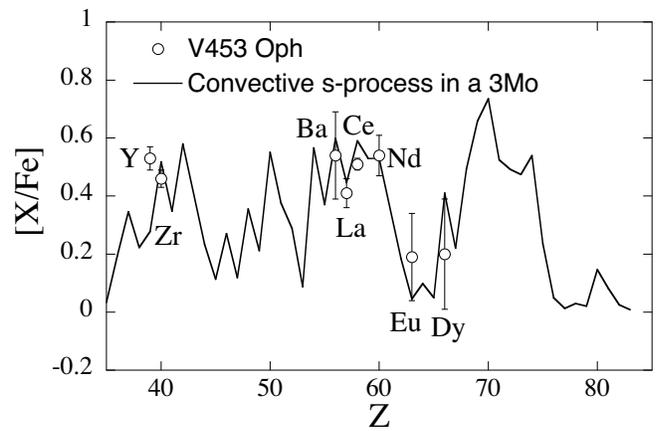}
\caption{ Comparison of the observed surface abundances (enrichment
of V453 Oph with respect to DS Aqr) with the s-process abundances
obtained within the convective s-process model in a  3\msun\
Z=0.0001 model star.}
  \label{fig_conv}
\end{figure}

As shown by \citep{Deroo2005}, there is however one s-process model
that could explain to some extent at least part of the
observations. This is the so-called convective s-process model that
can possibly develop at the base of hot thermal pulses in massive AGB
stars. Assuming again the extrinsic enrichment scenario to prevail, we
can imagine that a massive AGB companion star could have polluted
\object{V453 Oph} by stellar winds. In massive AGB stars, the
temperature at the base of the thermal pulse is hot enough for the
$^{22}$Ne($\alpha$,n)$^{25}$Mg to become active and be responsible for
some s-process nucleosynthesis. The major advantage of this scenario
in comparison with the PMP model is that the s-process enrichment in
the pulse can reach values similar to or even greater than the C
one. This is due to the large mass extent on which s-process seed
nuclei are irradiated by neutrons ($\rm{M}_{\rm pulse} \sim
10^{-3}$\msun) compared with the $10^{-5}$\msun resulting from the PMP
models. The resulting surface overabundances are shown in
Fig.~\ref{fig_conv}. Such a dilution gives rise to the corresponding
overabundances [C/Fe]=0.1, [N/Fe]=0.6 and [Mg/Fe]=0.4, and no Pb
overproduction contrary to the PMP nucleosynthesis. There is however a
drawback to this scenario: to reduce the C pollution, the star should
not have endured the third dredge-up episodes. It remains unclear what
star may or may not be subject to the third dredge-up, different
convective models with or without extra mixing mechanisms leading to
significantly different predictions. The presence of a binary
companion is an additional source of complexity.

Only in this very specific model and if no 3DUP
occurs to enrich the envelope in carbon, the surface enrichment of
s-elements can be reconciled with a low C-content. It is clear that,
although the comparison of the calculated and observed abundances is
quite satisfactory, the 3\msun\ Z=0.0001 model star can only have been
the putative companion to \object{V453 Oph}. Such a massive object of
low metallicity must have evolved into a white dwarf by now.

\section{Conclusion}\label{sect_conclusion} 

In this paper, an abundance analysis is presented of two very similar
RV\,Tauri stars of class C, \object{V453\,Oph} and
\object{DS\,Aqr}. Both objects are genuine post-AGB stars, although
both have no detected IR excess due to circumstellar
dust. These stars differ significantly when we look at their
photospheric abundance pattern. While \object{V453\,Oph} shows an
excess of light and heavy s-process elements,
\object{DS\,Aqr} shows neither. This conclusion is clearly
demonstrated by the relative analysis that was performed between both
objects.

The abundances derived for \object{V453\,Oph} are of particular
nucleosynthesis interest, since for the first time, the observed
overabundance of a s-process signature is not accompanied by a 
simultaneous C-enrichment. Moreover, the N abundance upper limit
excludes possible CN cycle destruction of the dredged-up carbon.  The
most favoured AGB s-process model based on the partial mixing of
protons at the time of the third dredge-up is inevitably accompanied
by a larger surface enrichment in C and therefore fails to account for the
present observation.

We explored three evolution scenarios to explain the abundance
distribution of \object{V453 Oph}. All have specific shortcomings.
The first possibility is that the parental cloud was initially
polluted by heavy elements.  The abundances observed in non-evolved
objects of the same metallicity ([Fe/H]$\sim -2.2$) show quite some
spread, illustrating that the Galaxy was not well mixed at these
epochs. Although some of these objects do exhibit significant heavy
element overabundances, a comparison with the abundances observed in
\object{V453 Oph} indicate that its parental cloud must have been
quite extremely polluted in r- but also s-elements.

Alternatively, the star may have been enriched during its evolution
either by mass transfer or internally. If the star is indeed
intrinsically enriched, this would suggest that low-mass objects of
low metallicity can be subject to s-process nucleosynthesis without
simultaneously polluting their envelope with C. At the present time,
it is impossible to reconcile such conditions with our present
understanding of AGB stars. If the star belongs to a binary system,
the companion may have polluted the actual post-AGB star but here
again the same C deficiency is difficult to reconcile with the
traditional s-process scenarios.  Moreover, there is still no
evidence of a WD companion.  Nevertheless, we explored the
possibility for a 3\msun\ Z=0.0001 model star to produce s-elements
within the thermal pulses only and conclude that such a convective
s-process model could to some extent reconcile the observed s-process
enrichment without significant C-enrichment.  Clearly, more
in-depth theoretical and observational studies are needed to further
understand if such a scenario could indeed be responsible for the
extraordinary chemical composition of the \object{V453 Oph} surface.

\bibliographystyle{aa}

\end{document}